# Dipole-field-assisted charge extraction in metal-perovskite-metal back-contact solar cells

Xiongfeng Lin[1]†, Askhat N. Jumabekov[2]†, Niraj N. Lal[1,2], Alexander R. Pascoe[1], Daniel E. Gómez[3,2], Noel W. Duffy[4], Anthony S. R. Chesman[2], Kallista Sears[2], Maxime Fournier[5], Yupeng Zhang[1], Qiaoliang Bao[1], Yibing Cheng[1,5], Leone Spiccia[6], Udo Bach[1,2,5,7]*

[1]Department of Materials Science & Engineering, Monash University, Clayton, Victoria 3800, Australia.

[2]CSIRO Manufacturing, Clayton, Victoria 3168, Australia.

[3]Plasmonics & Photochemistry laboratory, RMIT University, Melbourne, Victoria 3000, Australia.

[4]CSIRO Energy, Clayton, Victoria 3168, Australia.

[5]ARC Centre of Excellence in Exciton Science, Monash University, Clayton, Victoria 3800, Australia

[6]School of Chemistry, Monash University, Clayton, Victoria 3800, Australia.

[7]Melbourne Centre for Nanofabrication, Clayton, Victoria 3800, Australia

*Correspondence to: udo.bach@monash.edu.

†These authors contributed equally to this work.

**Abstract**

Hybrid organic-inorganic halide perovskites are low-cost solution-processable solar cell materials with photovoltaic properties that rival those of crystalline silicon. The perovskite films are typically sandwiched between thin layers of hole and electron transport materials, which efficiently extract photogenerated charges. This affords high-energy conversion efficiencies but results in significant performance and fabrication challenges. Herein we present a simple charge transport layer-free perovskite solar cell (PSC), comprising only a perovskite layer with two interdigitated gold back-contacts. Charge extraction is achieved via self-assembled molecular monolayers (SAMs) and their associated dipole fields at the metal/perovskite interface. Photovoltages of approximately 600 mV generated by SAM-modified PSCs are equivalent to the built-in potential generated by individual dipole layers. Efficient charge extraction results in photocurrents of up to 12.1 mA cm$^{-2}$ under simulated sunlight, despite a large electrode spacing.

**Introduction**

Hybrid organic-inorganic halide perovskites are amongst the most promising materials for photovoltaic applications[1-4]. Based on earth abundant, solution-processable, low-cost materials, this subclass of perovskite compounds exhibit attributes that challenge silicon's predominance in photovoltaics. Efficient charge separation in PSCs is typically achieved by carrier selective charge transport layers (CTLs), which exist as either conventional transport materials or ultrathin work function modifiers[5]. The fabrication of pinhole and defect free multi-layer structures is challenging, especially for moisture-sensitive solution-processable materials. Furthermore, key challenges facing the field of PSCs, such as enhanced recombination, poor stability, and hysteresis during current-voltage scans are linked to carrier selective charge transport layers (CTLs) and their respective CTL/perovskite interfaces[6-8].

The photovoltaic properties of CTL-free PSCs, in which metals or graphene directly contact the perovskite to create Schottky junction solar cells[9-14], depend strongly on the built-in potential established by the work function difference of the two contacts (Fig. 1a). These differences can be temporarily induced in gold/perovskite/gold solar cells by applying a large external bias (>2 V) across the symmetric contacts (poling), resulting in migration and accumulation of $CH_3NH_3^+$ and $I^-$ ions at the respective contacts[13-20]. However, poling is not sufficient to provide a stabilized power output.

Persistent work function modifications of metals and semiconductors can be induced by self-assembled monolayers (SAMs) featuring oriented molecules with permanent dipole moments[21-24]. The built-in potential of heterojunctions can be adjusted accordingly by placing these SAMs directly at the junction interface. This concept has been demonstrated successfully in organic light-emitting-diodes (OLEDs) and organic field-effect transistors (OFETs), but not explicitly within solar cells[25-27].

Herein we present a simple gold/perovskite/gold back-contact PSC (bc-PSCs) with a built-in potential and photovoltaic response owing exclusively to dipole SAMs present at the gold/perovskite interfaces (Supplementary Fig. 1). An interdigitated gold microelectrode array (IDA) on glass forms the back-contact of the solar cell (Supplementary Fig. 2). The SAMs are formed by two para-substituted thiolbenzene derivatives with opposing molecular dipoles (Fig. 1b). Both molecules are insulators and do not play an active role in the charge transport process. We present a method to selectively modify the two sets of IDA microelectrodes, termed 'a' and 'b', with the SAM compounds. This modification induces a work function difference between the two microelectrodes of up to 600 mV, as observed through Kelvin probe force microscopy (KPFM). Bc-PSCs are fabricated by spin-coating a perovskite precursor solution onto the SAM-modified IDA in a single step. The open-circuit potential ($V_{OC}$) of these bc-PSCs is shown to be equivalent to the work function difference induced by the SAM modification ($\phi_m^{(b)} - \phi_m^{(a)}$). Photocurrent mapping measurements reveal near uniform (<25%) charge collection across the cell, giving rise to short-circuit photocurrent densities ($J_{SC}$) of up to 12.1 mA cm$^{-2}$ (Supplementary Fig. 3) despite the large 6.5 µm center-to-center (distance from the center of electrode 'a' to the center of adjacent electrode 'b') and 4.3 µm edge-to-edge (distance from the edge of electrode 'a' to the edge of adjacent electrode 'b') electrode distances.

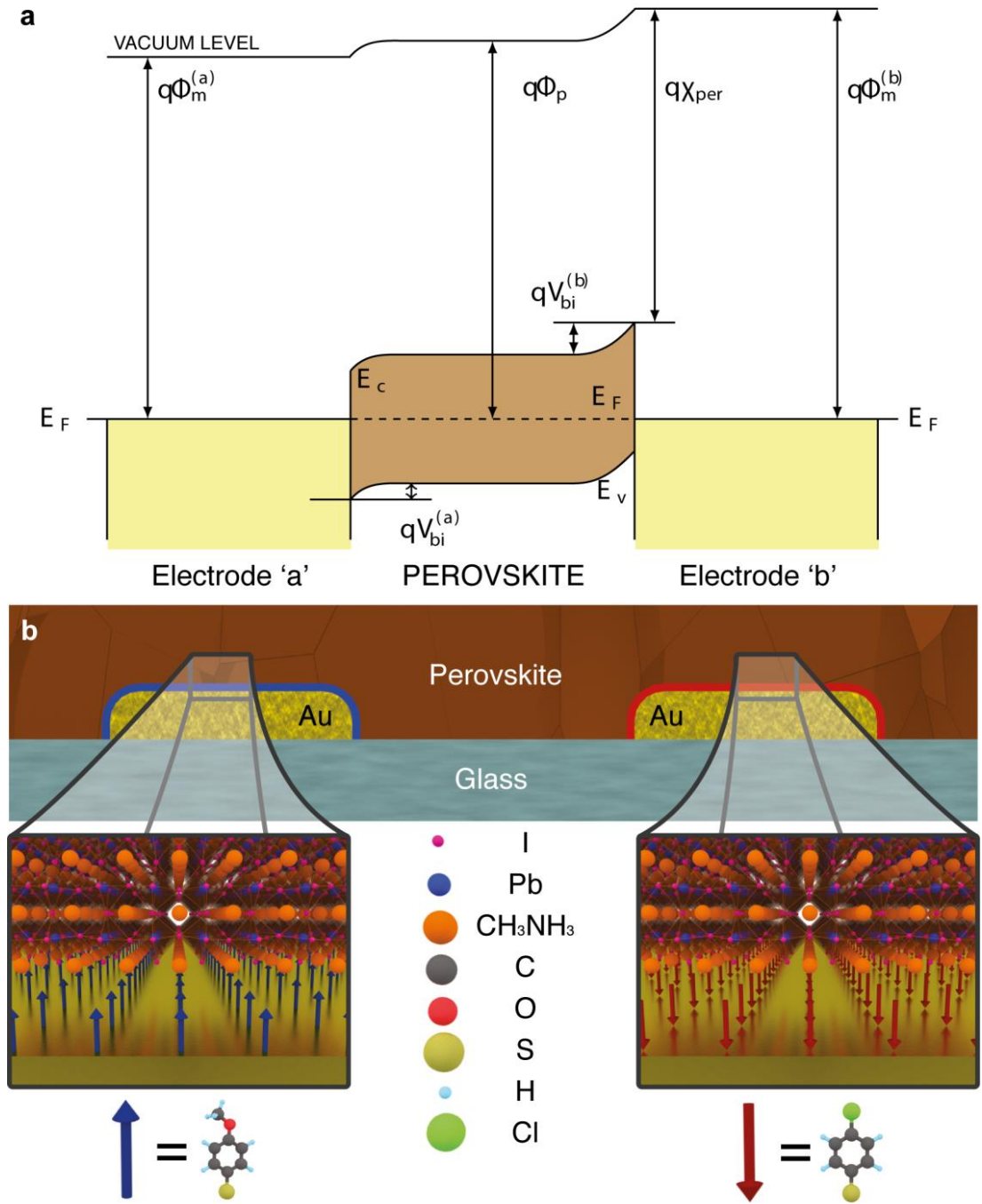

**Figure 1 | Back-contact metal/perovskite/metal solar cells.** (**a**) Schematic energy band diagram of a metal/perovskite/metal solar cell at thermal equilibrium in the dark. The work function of the MAPbI$_3$ perovskite ($q\phi_p$) is situated between those of the metal electrodes 'a' ($q\phi_m^{(a)}$) and 'b' ($q\phi_m^{(b)}$). The overall built-in potential of the solar cell is equal to the sum of the built-in potentials at the metal/perovskite contacts. $q\chi_{per}$ corresponds to the electron affinity of MAPbI$_3$. (**b**) Cross-section diagram of a dipole SAM modified back-contact gold/MAPbI$_3$/gold solar cell. Electrode 'a' (anode, left) is modified with a molecular monolayer of 4-methoxythiophenol (OMeTP) with a molecular dipole of −2.67D. Electrode 'b' (cathode, right) is modified with monolayer of 4-chlorothiophenol (ClTP) with a molecular dipole of +1.41D.

## Results

**Work function modification of IDA.** The detailed steps used to fabricate the modified interdigitated microelectrode arrays are given in the Methods section (see Supplementary Fig. 4). In brief, the IDA is immersed into an OMeTP solution, resulting in the formation of OMeTP SAMs on both microelectrodes 'a' and 'b'. The OMeTP SAM on electrode 'b' is subsequently desorbed electrochemically. The complete IDA is then exposed briefly to a solution of ClTP in order to form the respective SAM on microelectrode 'b'.

KPFM was used to monitor the surface potential changes of the gold IDAs after each fabrication step. KPFM allows for the direct mapping of the contact potential difference ($CPD = \phi_{sample} - \phi_{tip}$) between the AFM tip and the sample surface, hence quantifying the work function difference ($\phi_m^{(b)} - \phi_m^{(a)}$) between electrodes 'b' and 'a'. Fig. 2a shows the KPFM image of an IDA section prior to SAM modification. All fingers initially have a similar *CPD* of approximately +100 mV. Fig. 2c shows the KPFM image after OMeTP adsorption and electrochemical desorption from electrode 'b'. The formation of the OMeTP SAM on electrode 'a' is confirmed indirectly by the 400 mV *CPD* increase relative to the bare electrode, while electrode 'b' increases by approximately 100 meV due to the incomplete desorption of the OMeTP SAM. The subsequent exposure to the C1TP solution further increases this work function difference to 550 – 580 meV. These *CPD* results agree well with photoelectron spectroscopy in air (PESA) measurements on SAM-modified gold films (Supplementary Fig. 5). *CPD*, PESA and Raman spectroscopy measurements provide evidence for the incomplete electrochemical desorption of OMeTP from electrode 'b' and some co-adsorption of ClTP on electrode 'a' via thiol-exchange reactions during the final modification step (Supplementary Fig. 6). These measurements indicate that stronger built-in potentials could be obtained by avoiding these cross-contamination processes.

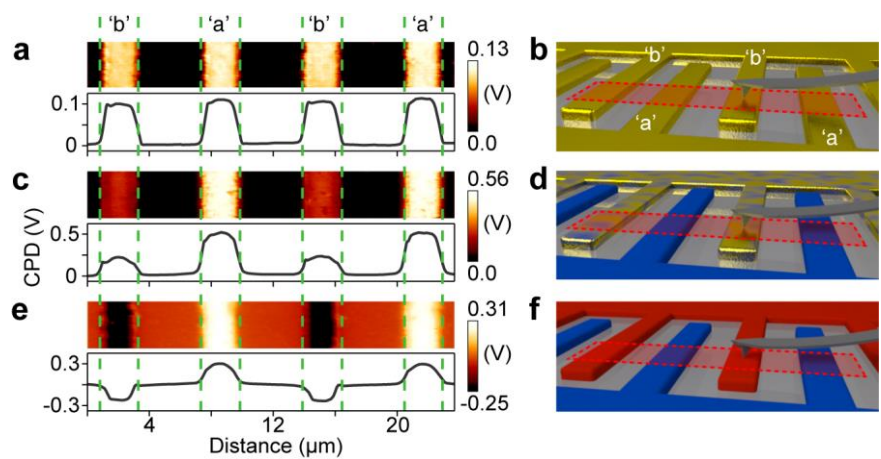

**Figure 2 | Kelvin-probe-microscopy of IDAs during molecular modification.** Contact potential difference maps across sets of interdigitated fingers and illustrations of the KPFM imaging experiment for an unmodified IDA (**a**+**b**), an IDA after exposure to OMeTP and subsequent electrochemical desorption from electrode 'b' (**c**+**d**) and an IDA after the final modification step with ClTP (**e**+**f**).

**Performance of dipole modified bc-PSCs.** We fabricated 'a'-MAPbI$_3$-'b' bc-PSCs by spin-coating a perovskite precursor solution containing PbI$_2$ and CH$_3$NH$_3$I in dimethyl sulfoxide (DMSO) and N-methyl-2-pyrrolidone (NMP) (v/v 7:3) onto the OMeTP and ClTP SAM-modified IDAs (see Methods for details). The photovoltaic performance was measured via photocurrent density versus voltage (*J-V*) curves under simulated sunlight (AM 1.5 G; 1000 W m$^{-2}$). The active area of the device is defined by the interdigitated area of the gold electrodes, which is 2 × 2 mm (Supplementary Fig. 2). Fig. 3a shows *J-V* curves for a typical device recorded in forward scan mode (0 V → $V_{OC}$) and reverse scan mode ($V_{OC}$ → 0 V). The scans reveal negligible hysteresis and a $V_{OC}$ of 560 and 570 mV for forward and reverse scan modes, respectively. This is similar to the work function difference of 550–580 meV measured for the SAM-modified back-contact plane. A strong correlation between $V_{OC}$ and the work function difference between anode and cathode was also observed for bc-PSCs where only one of the two electrodes was modified with either a OMeTP or ClTP SAM (see Supplementary Fig. 7 and Fig. 8). This finding suggests that larger $V_{OC}$ values can be achieved using SAMs that further increase the work function difference between the contacts. Bc-PSCs fabricated without contact modification produced devices with a $V_{OC}$ of less than 100 mV, confirming that the built-in potential was fundamentally a result of the SAMs modification. The recorded $J_{SC}$ values of 11.4 mA cm$^{-2}$ and 11.3 mA cm$^{-2}$ for forward and reverse scan modes, respectively, are high, considering the large center-to-center (6.5 µm) and edge-to-edge (4.3 µm) electrode distances. With fill factors (*FF*) of 40.5% and 40.1%, power conversion efficiencies (*PCE*) of 2.59% and 2.57% were obtained for the unencapsulated device in forward and reverse scan modes, respectively (see Fig. 3a).

The *J-V* curves after encapsulation (Supplementary Fig. 2d) show a slight decrease in the photovoltaic parameters for the cell (Fig. 3a). The $V_{OC}$, $J_{SC}$ and *FF* of the encapsulated device are 0.55V, 9.73 mA cm$^{-2}$ and 36.7%, respectively, with a PCE of 1.96% for reverse scan mode. In forward scan mode the encapsulated device shows 0.54 V, 9.87 mA cm$^{-2}$, and 36.7% for the $V_{OC}$, $J_{SC}$ and *FF*, respectively, resulting in a PCE of 1.96%. Slight decreases of photovoltaic performance may originate from reflective losses due to the introduction of the glass cover over the perovskite layer, as well as overall device degradation due to the encapsulation process. The external quantum efficiency (EQE) of an encapsulated device was seen to be largely independent of wavelength in the 350–900 nm range, with a value in excess of 33%. The determined AM 1.5G short-circuit current obtained from the EQE spectrum (red curve in Fig. 3b) of 9 mA cm$^{-2}$ is in agreement with the $J_{SC}$ value of 9.73 mA cm$^{-2}$ derived from the *J-V* characteristics of the encapsulated device (Fig. 3a). Monitoring the performance of SAM modified bc-PSCs at the maximum power point (MPP) as a function of time revealed that the solar cells exhibit a stable power output ($P_{MPP}$) (Fig. 3, c and d), with a value of ~2.3%. The highest performing bc-PSC fabricated in this study showed an efficiency of 3.11% (see Supplementary Fig. 3 and Fig. 8a) with the $J_{SC}$ value of 12.1 mA cm$^{-2}$. For comparison, *J-V* measurements on unmodified Au-MAPbI$_3$-Au devices showed no photovoltaic effect (Supplementary Fig. 8b)[13,15,16]. Poling induced by a 2 V external bias for 80 seconds (Fig. 3, c and d) initially resulted in a $P_{MPP}$ of 0.13 mW cm$^{-2}$, however this photovoltaic effect was momentary, with the $P_{MPP}$ dropping by approximately 90% over the 150 s observation period.

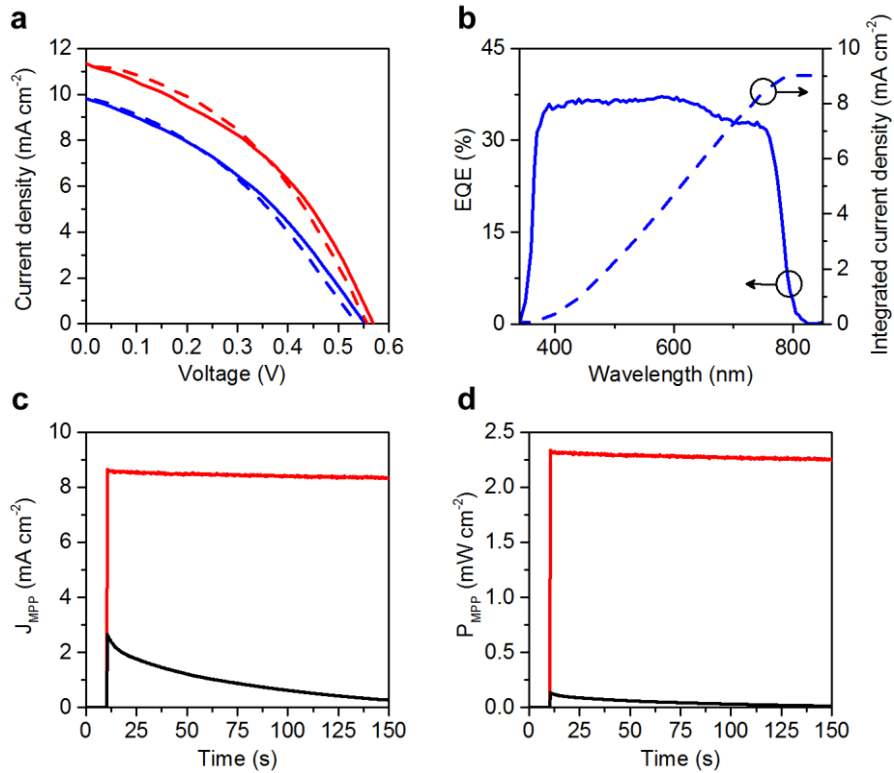

**Figure 3 | Photovoltaic properties of SAM-modified bc-PSCs.** (**a**) *J-V* curves of a dipole-modified gold-perovskite-gold bc-PSC measured under standard solar irradiation (AM 1.5G, 1000 W m$^{-2}$), reverse scan (solid blue, encapsulated; solid red, unencapsulated) and forward scan (dashed blue, encapsulated, dashed red, unencapsulated). (**b**) Spectral response (EQE, solid line) and the corresponding integrated short-circuit current density (dashed line) of a SAM-modified PSC. (**c**) Time evolution of the maximum power point photocurrent density ($J_{MPP}$) of a SAM-modified (red) and poled, unmodified bc-PSC (black) and their power output (**d**).

**Photocurrent and photoluminescence mapping.** Photocurrent (PC) and photoluminescence (PL) maps were recorded simultaneously using a confocal laser beam scan across the solar cell's active area (Fig. 4, a and b, see Methods for details). The PC profile features photocurrent minima above both collection electrodes and maxima within the electrode gaps, consistent with previous reports of balanced electron and hole diffusion lengths in MAPbI$_3$[28]. For charges generated at the mid-point between 'a' and 'b' electrodes, the minimum travel distance for collection is approximately 2 µm, while one of the two carriers is required to travel at least 5.3 µm when generated directly above an electrode. The PC profile shows the overall variation in EQE is less than 25% across the entire cell, indicating excellent charge-carrier extraction. Variations in the perovskite film thickness as source of PC or PL fluctuations can be excluded, with >90% of the scanning laser light absorbed within the top 200 nm of the MAPbI$_3$ film.

The PL map shows luminescence maxima for excitation above OMeTP modified 'a' electrodes and minima above ClTB modified 'b'-electrodes. The luminescence intensity above both electrodes differs by about two orders of magnitude (Supplementary Fig. 9). The PL appears mostly featureless above electrodes 'b' but shows distinct grain-like features near the center of each electrode 'a' and inside the electrode gaps. This is consistent with the AFM topography image (Fig. 4d and Fig. 4e) that reveals large multi-micron-sized MAPbI$_3$ crystalline regions along electrode 'b', as well as a pronounced microgap located at the center of each 'a' electrode. This provides some indication that the MAPbI$_3$ film morphology appears to be the major contributing factor to the PL variations observed. Isolated, highly luminescent MAPbI$_3$ crystalline regions within this microgap are a likely source for the strongly luminescent features observed in the PL map. The observed topography patterns result from the specific bc-PSC architecture and the different wetting properties of the OMeTP- and ClTB-functionalized electrodes affecting the crystallization and film formation process (Supplementary Fig. 10), as confirmed by contact angle measurements on bare and SAM-modified gold films, in which the OMeTP-modified gold films showed significantly better wetting than the bare or ClTB-modified gold films (Supplementary Fig. 11).

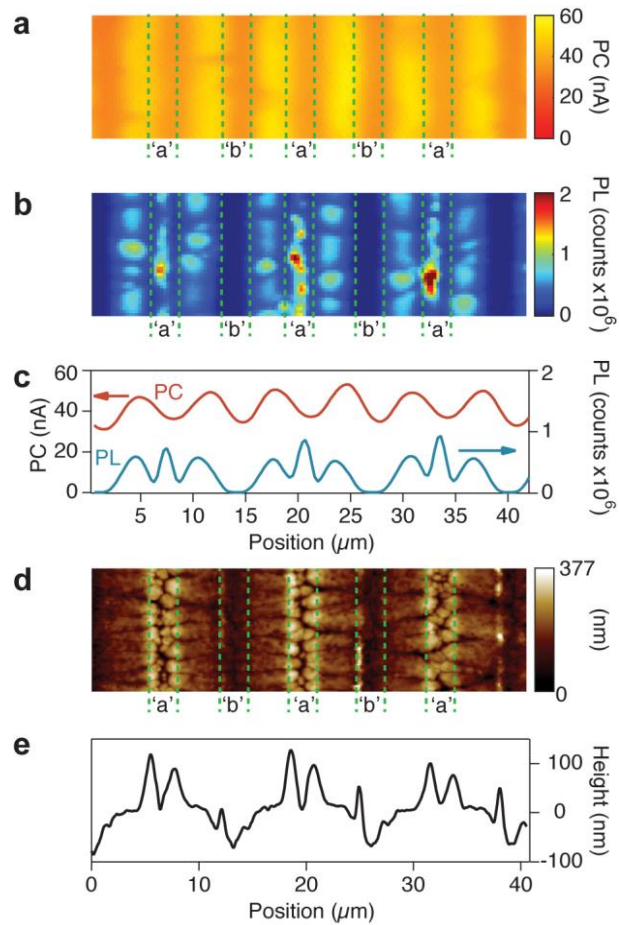

**Figure 4 | Spatial photocurrent (PC) and photoluminescence (PL) mapping of SAM-modified bc-PSC.** (**a**) Spatial PC map of a SAM-modified bc-PSC at short-circuit condition. (**b**) Spatial PL map of a SAM-modified bc-PSC at short-circuit condition. (**c**) PC and PL profile obtained by averaging the PC and PL maps along the vertical direction. (**d**) AFM topography map of a SAM-modified bc-PSC. (**e**) AFM topography map profile of a SAM-modified bc-PSC. The scanned area in all subpanels is 42 × 12 μm.

**Modelling of theoretical PCE limit and charge collection efficiency.** A number of simulations were carried out in order to gage the performance limit of Schottky junction PSCs and more specifically back-contact Schottky PSCs. First, we estimated the maximum theoretical PCE for a MAPbI$_3$ Schottky junction solar cell to be 25%, based on a thermionic emission model[29] (Fig. 5a). This efficiency limit is expected for an ideal Schottky junction

MAPbI$_3$ solar cell in which the junction barrier height equals the value of its bandgap of 1.6 eV. Similar efficiency limits of 25-27% have recently been derived for PSCs with conventional p-i-n heterojunction structures[30]. For the OMeTP and ClTP-modified bc-PSCs presented here we calculated a barrier height of 1.2 eV and a corresponding maximum theoretical efficiency limit of 15% (see Methods for details). Further transfer matrix optical modelling of the 'a'-MAPbI$_3$-'b' device yielded an upper limit of 19.9 mA cm$^{-2}$ as the maximum theoretical $J_{SC}$ under AM1.5G solar irradiation for the bc-PSCs reported here. This estimation is simply based on the overall optical light harvesting efficiency of the device (see Supplementary Fig. 12 and Methods for details)[31].

Ultimately, the charge collection efficiency of bc-PSCs is dependent on the relationship between electrode separation distance and the charge carrier diffusion length $L_d$. Fig. 5b depicts the collection efficiency (normalised $J_{SC}$) as a function of $L_d$ and electrode separation distance, and is calculated based on the formalism of Taretto et al.[32] with a built in voltage of 0.6 V and balanced diffusion lengths. The model predicts that charge collection efficiencies close to 100% for situations in which the electrode spacing is comparable to, or smaller than, the diffusion length. We also used the same formalism by Taretto et al.[32] (see Methods and Supplementary Fig. 13 for details) to calculate a charge carrier diffusion length of 350 nm for the encapsulated 'a'-MAPbI$_3$-'b' bc-PSCs, based on the photocurrent mapping data shown in Fig. 4c[32-34]. Based on this value and the geometrical electrode spacing (edge-to-edge) of 4.3 µm the model predicts a charge carrier collection efficiency for the current configuration of approximately 40% (see dashed lines in Fig. 5b). This is in excellent agreement with the EQE data (Fig. 3b), which also shows around 40% for the same encapsulated 'a'-MAPbI$_3$-'b' bc-PSCs at illumination wavelength of 532 nm.

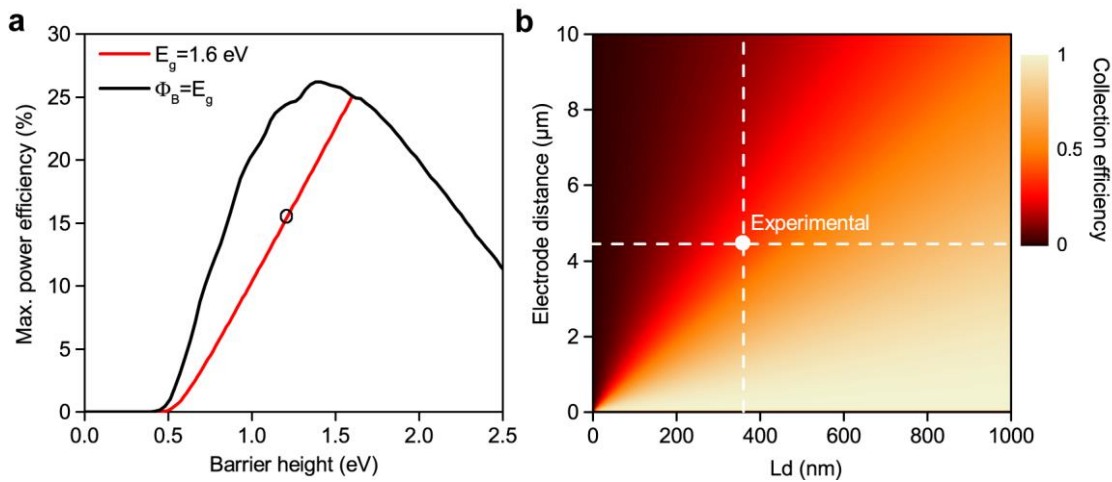

**Figure 5 | Modelling of theoretical PCE limit and collection efficiency.** (**a**) Theoretical power conversion efficiency limit (red) as a function of barrier height $\Phi_B$, calculated for a Schottky junction MAPbI$_3$ bc-PSC under AM1.5G irradiation, according to Pulfrey et al.[29] (green). The temperature of the device was assumed to be $T$=300 K, the effective Richardson constant (thermionic emission) $A^{**}$ = 120 A cm$^{-2}$ K$^{-2}$. The black curve shows the maximum power conversion efficiency for the ideal situation in which the barrier height is equal to the band-gap. The black circle indicate the maximum theoretical PCE achievable for OMeTP and ClTP-modified 'a'-MAPbI$_3$-'b' bc-PSC. (**b**) Contour map of charge collection efficiency as a function of diffusion length $L_d$ and electrode separation distance. The dashed lines indicate the current experimental device parameters.

**Discussion**

In conclusion, we have presented the first implementation of a photovoltaic device in which charge extraction is solely achieved by a dipole field generated via molecular monolayers. We expect this general concept to be adaptable to a wide range of absorber, electrode and molecular dipole materials well beyond the capabilities shown here. Through modelling we demonstrated that high-performance solar cell devices can be obtained using back-contact electrode design with a higher work function asymmetry between the electrodes and optimised electrode spacing. A number of important technical advantages arise from abolishing the need for additional charge transport layers in concert with back-contact designs. Novel back-contact design concepts[35] combined with scalable low-cost nanofabrication techniques[36] offer the potential for boosting solar cell efficiencies while providing scope for mass-production at low cost.

**Methods**

**Chemicals.** Unless specified otherwise, all materials for perovskite precursor preparation were purchased from either Alfa Aesar or Sigma-Aldrich and used as received. 4-Methoxythiophenol (OMeTP, Sigma-Aldrich, ≥98.0%) and 4-chlorothiophenol (ClTP, Apollo Scientific Limited) were used without further purification. Reverse osmosis (RO) purified water with a quoted resistivity of 1 MΩ cm at 25 °C was used for all desorption experiments.

**Perovskite Precursor Preparation.** Methylammonium iodide, $CH_3NH_3I$ (MAI) was synthesized according to a previous study[37]. To synthesize MAI, 24 mL methylamine ($CH_3NH_2$, 33% in ethanol) was mixed with 10 mL hydroiodic acid (HI, 57% in water) in 100 mL ethanol. The solvent was removed under vacuum to give a powder of MAI, which was subsequently washed with diethyl ether and used without further purification.

**Fabrication of Gold Interdigitated Microelectrode Arrays (IDAs).** Initially, 25 × 25 mm soda-lime glass substrates were cleaned by sequential sonication steps in acetone and 2-propanol solutions, each performed for 15 min. After drying with compressed air, a 2 μm thick photoresist layer was spin coated onto the glass substrates and baked at 110 °C for 2 min. The substrates were then exposed to UV light through a chrome photo-mask with the desired pattern. In the developmental step, the UV-exposed sections of the photoresist layer were removed by washing the substrate in a developer solution, resulting in a portion of the substrates being protected by the photoresist mask. In the next step, 10 nm Ti and 60 nm Au layers were sequentially evaporated onto the substrates using an electronbeam evaporator. Finally, the remainder of the polymer mask was removed by sonicating the substrates in acetone (lift-off process), with six IDA electrodes deposited onto a single substrate (Supplementary Fig. 2a). The yield of the IDA fabrication process was ~98%. The interdigitated area of the electrodes is 2 × 2 mm, in which the width of the IDA fingers is ~2.2 μm and the gap between the interdigitated fingers is ~4.3 μm (Supplementary Fig. 2b).

**Work Function Modification of Gold IDAs with Thiophenol-Based Self-Assembled Monolayers (SAMs).** The selective deposition of 4-methoxythiophenol (OMeTP) and 4-chlorothiophenol (ClTP) onto the separate gold components of the IDA requires a stepwise protocol (Supplementary Fig. 4a). Prior to modification, the gold IDA was treated with ozone plasma (Harrick Plasma) for 18 min. For OMeTP-modification, the substrate was immersed in a 20 mM OMeTP in ethanol solution for 24 hours in an inert atmosphere. The OMeTP-modifed IDA was then rinsed with anhydrous ethanol and dried under a nitrogen flow. Desorption of the OMeTP monolayer from one half of the IDA was achieved through an electrochemical process (see below), followed by rinsing of the substrate with RO water and ethanol, and drying with a nitrogen flow. Desorption of the OMeTP monolayer on one half of the IDA was verified by kelvin probe force microscopy (KPFM) measurements (see below). A ClTP monolayer was then deposited onto the unmodified part of the IDA by soaking the substrate in a 10 mM ClTP in ethanol solution for 1 min. The OMeTP- (electrode 'a') and ClTP-modified (electrode 'b') IDA was then rinsed with anhydrous ethanol and dried under a nitrogen flow.

**Electrochemical Treatment for Desorption of OMeTP.** Desorption of the OMeTP monolayer from one half of the IDA was obtained by using a Bio-Logic VSP instrument configured in bipotentiostat mode (Supplementary Fig. 4b)[38-40]. The working electrodes were the two halves of the IDA. The counter-electrode was tin wire and the reference electrode was Ag|AgCl|NaCl (3 M) (CH-Instrument). All experiments were carried out in ambient air at 21 °C in a 0.5 M KOH aqueous solution. No degassing process was applied to the electrolyte during the desorption procedure.

A potential of −0.144 V vs. Ag/AgCl (corresponding to the open circuit potential of the modified surface) was constantly maintained for the upper half of the IDA (see Supplementary Fig. 4a) in order to prevent any desorption of OMeTP. At the same time, −1.05 V vs. Ag/AgCl was applied to the lower half of the IDA until the evolution of current over time stagnated over about 200 µA, corresponding to the capacitive current of an unmodified gold surface (Supplementary Fig. 4b).

**Perovskite Deposition.** $PbI_2$ and MAI were dissolved in a 1:1 molar ratio in a solution of dimethyl sulfoxide (DMSO) and N-methyl-2-pyrrolidone (NMP) (v/v 7:3) to obtain a 1.74 M perovskite precursor solution. The $MAPbI_3$ thin-films were formed in an inert atmosphere by spin coating the precursor solution onto the substrates using a two-step program (500 rpm for 10 s, then 6000 rpm for 50 s). A $N_2$ gas stream was introduced at the 20 s mark of the second step, and sustained for a further 20 s. The films were then transferred into a covered glass petri dish containing two droplets (0.5 µL each) of a DMSO/NMP (v/v 7:3) solution. Perovskite films were solvent annealed at 100 °C inside a closed petri dish for 5 min, and the petri dish was then opened for another 5 min of annealing. The film thickness on top of the IDA electrodes has an average value of ~340 nm. However, due to the difference in wettability of anode and cathode ('a' and 'b', respectively) the film thickness over the electrode 'a' is ~240 nm, while over the electrode 'b' it is ~440 nm (Supplementary Fig. 10).

**Kelvin Probe Force Microscopy (KPFM) and Atomic Force Microscopy (AFM).** KPFM and AFM measurements were performed on a Dimension Icon (Veeco) in air. Chromium-platinum coated conductive probes (ElectriMulti75-G, BudgetSensors) were used for the measurements. The scanning area and scanning rate were 25 × 3.125 µm and 0.2 Hz, respectively. The surface roughness has very little influence on the recorded *CPD* values (see Supplementary Fig. 14), whereas the lift height of the AFM tip can greatly affect the magnitude of the *CPD* (see Supplementary Fig. 15). The lift height was 15 nm for all KPFM measurements.

**Raman Spectrum on Fully Modified Interdigitated Electrode.** Raman spectroscopy measurements were performed on a Renishaw inVia confocal microscope with a 785 nm laser operating at 100 % power, in the range 250–1800 nm with a resolution of 2 nm. Acquisition time was 10 seconds per scan with the final spectra from the co-addition of 40 scans. Prior to thiol deposition, gold IDAs were subjected to a surface roughening step via an electrochemical process in order to induce a surface-enhanced Raman scattering (SERS) effect[41]. In the absence of such a step the thiols could not be detected on the surface of the IDAs by Raman spectroscopy.

**Photoelectron Spectroscopy in Air (PESA).** The work functions of thiol modified gold surfaces were investigated using a Riken Keiki Photoelectron Spectrometer (Model AC-2) at 20 nW.

**Solar Cell Characterization.** The current density-voltage (*J-V*) characteristics of devices were measured in an inert atmosphere using a Keithley 2400 source meter under 1.5 G light illumination from a solar simulator (Newport 150 W Xenon lamp coupled with an AM 1.5 G solar spectrum filter). The light intensity was calibrated to 100 mW cm$^{-2}$ using a secondary reference photodiode (Hamamatsu S1133, with KG-5 filter, 2.8 × 2.4 mm photosensitive area), which was calibrated by a certified reference cell (PVMeasurements, certified by NREL). The cells were measured from the front side (absorber side) with a back reflector made of Al foil. For comparison, the *J-V* characteristics for back side (glass side) illumination were also recorded (see Supplementary Fig. 16). The devices were stored in an inert atmosphere and *J-V* characteristics were obtained after a prolonged storage time (see Supplementary Fig. 17). All the *J-V* characteristics were recorded without a mask since the collected photocurrent originated only from the interdigitated area (2 × 2 mm) of the cell (see Supplementary Fig. 18).

**Photocurrent (PC) and Photoluminescence (PL) Mapping.** Photocurrent and PL mapping measurements are carried out using a combined detection system, based on a WITec Alpha 300R confocal microscope. A 532-nm laser is fiber-coupled (Thorlabs P5-405BPM-FC-2) through a fiber bench (Thorlabs FBP-A-FC), optically chopped (C995, Terahertz Technologies, Inc., 572 Hz) and coupled into the microscope, with the resulting 1 μm diameter spot size and 2.8 μW power measured at the sample. PL is recorded at short-circuit and summed between 720 and 780 nm, and PC is recorded at short-circuit current using a preamplifier (FEMTO DLPCA-200, Messtechnik GmbH) and a lock-in amplifier (FEMTO LIA-MV-150). The devices were illuminated from the front side (absorber side) and measurements were carried out in an ambient atmosphere with encapsulated devices.

**External Quantum Efficiency (EQE).** EQE data was collected using an Oriel 150W Xe lamp coupled to a monochromator and an optical fibre in an ambient atmosphere. The devices were illuminated from the front side (absorber side). The light source was chopped at 27 Hz and the electrical signal was collected under short circuit conditions using a low noise current pre-amplifier (SR570, Stanford Research Systems) and lock-in amplifier (SR830 DSP, Stanford Research Systems). A standard, filtered Si cell from Peccell Limited, which was cross-calibrated with a standard reference cell traceable to the National Renewable Energy Laboratory, was used as the reference. The devices were encapsulated to avoid severe degradation during measurement. However, some gradual degradation of the devices is observed during the measurements despite encapsulation.

**Contact Angle.** Contact angle measurements were obtained using a Contact Angle System OCA (Dataphysics). Milli-Q water was dropped onto samples with a dosing volume of 2 μL and a dosing rate of 1 μL s$^{-1}$.

**Scanning Electron Microscopy (SEM) and Focus Ion Beam (FIB).** Cross sections of the samples were prepared by FIB and the SEM cross section images were taken using a FEI Helios NanoLab600 Dual Beam FIB-SEM at a 5 kV accelerating voltage.

**Raman Spectrum of Fully Modified Interdigitated Electrode.** Raman spectroscopy confirmed the selective deposition of the functionalized thiols on finger regions of the gold IDAs (Supplementary Fig. 6), with the Raman spectra of the thiols matching literature values[42]. The fingers of the gold IDAs contained the same characteristic signals of the thiols as the connected bay regions (pad for contact), with the addition of a broad peak centered at 1400 cm$^{-1}$. This peak can be attributed to photoluminescence of the soda lime glass substrate caused by the 785 nm laser, which arises from the size of the laser beam being broader than the width of the IDA finger.

The region in which OMeTP was deposited also contained trace amounts of ClTP. This is evidenced by a shoulder on the large peak at 1591 cm$^{-1}$ and a broadening of the base of the peak at 1081 cm$^{-1}$, which are due to the single peak at 1568 cm$^{-1}$ and the three peaks centered at 1081 cm$^{-1}$, respectively, present in the spectrum of pure ClTP. This is either due to the partial exchange of OMeTP for ClTP on the gold surface during the second thiol deposition step, or ClTP depositing onto the gold surface in regions where the OMeTP monolayer deposition was incomplete.

**Transfer Matrix Optical and Collection Efficiency Modelling.** Transfer matrix optical modelling of the structure was undertaken in Matlab with bespoke software based on the formalism of Pettersson *et al.*[31]. The structure is modelled with 70 nm Au and 280 nm perovskite for the contact region (metal-semiconductor interface), and 350 nm of perovskite over the non-contact region (MAPbI$_3$-glass interface). An aluminium reflector is included at the rear of the optically-thick glass substrate, with reflection, absorption and transmission spectra weighted over the contact (width 2.5 μm) and non-contact (width 3.5 μm) regions. Perovskite *n*, *k* data are taken from Löper *et al.*, gold from Johnson *et al.*, and aluminium from Rakić *et al.*[43-45]. A photocurrent of 19.9 mA cm$^{-2}$ is predicted under incident AM 1.5G solar spectrum, assuming ideal collection efficiency, with >10% gain expected from the inclusion of a simple quarter-wavelength anti-reflection coating, consisting of 110 nm PMMA under a 350 nm thick perovskite absorber layer.[46]

The semi-empirical charge carrier diffusion length was obtained by fitting the experimental PC profile (Fig. 4c) using the formalism of Taretto *et al.*[32] (Supplementary Fig. 13a). The best fit in respect to peak-to-valley percentage variation of the PC profile (see Supplementary Fig. 13b) was chosen to be the $L_d$ value that corresponds to the experimental conditions presented here.

**Evaluation of the Energy Band Diagrams for Symmetric and Asymmetric Device Constructs at Thermal Equilibrium in the Dark.** For the construction of the band diagrams of the symmetric and asymmetric devices, the values of the band gap of MAPbI$_3$, $E_g$ =1.6 eV, the electron affinity of MAPbI$_3$, $q\chi_{per}$ =3.9 eV, and the work function of MAPbI$_3$, $q\Phi_s$ =4.7 eV, were taken from the literature[47,48]. The work functions of pristine gold (Au), OMeTP-modified ('a') and ClTP-modified ('b') gold were obtained from the PESA measurements (see below). For the symmetric device (Au-MAPbI$_3$-Au) the value of the work functions of both the 'left', $q\Phi_m^{(a)}$, and 'right', $q\Phi_m^{(b)}$, pristine gold contacts was ~4.9 eV, whereas in the asymmetric device ('a'-MAPbI$_3$-'b') the work function values for the OMeTP-modified 'left',

$q\Phi_m^{(a)}$, and ClTP-modifed 'right', $q\Phi_m^{(b)}$, gold contacts were ~4.5 eV and ~5.1 eV, respectively.

1) Symmetric device

In the symmetric Au-MAPbI$_3$-Au device the band diagram for the metal-semiconductor (M-S) Schottky interface at the 'left' contact was the same as for the 'right' contact (see Supplementary Fig. 1a), and formed back-to-back connected Schottky diodes. Therefore, it was sufficient to derive values for the barrier height, $q\Phi_{Bn}$, and the built-in potential, $qV_{bi} = qV_{bi}^{(a)} = qV_{bi}^{(b)}$, at the M-S interfaces for only one of the contacts, i.e. the 'left' contact, as follows[49,50]:

$$q\Phi_{Bn} = q(\Phi_m^{(a)} - \chi_{per}) = 4.9\,eV - 3.9\,eV = 1\,eV$$

$$qV_n = q(\Phi_s - \chi_{per}) = 4.7\,eV - 3.9\,eV = 0.8\,eV$$

$$qV_{bi} = q(\Phi_{Bn} - V_n) = q(\Phi_m^{(b)} - \Phi_s) = 0.2\,eV \quad \text{- 'upwards' band bending.}$$

For other symmetric devices such as Au-OMeTP-MAPbI$_3$-OMeTP-Au and Au-ClTP-MAPbI$_3$-ClTP-Au the values for $q\Phi_{Bn}$ and $qV_{bi}$ can be derived in a similar fashion.

2) Asymmetric device

In the asymmetric 'a'-MAPbI$_3$-'b' device the S-M Schottky interfaces at the 'left' and 'right' are not the same, due to different work functions of the 'left' and 'right' metal contacts ($\Phi_m^{(a)} \neq \Phi_m^{(b)}$) (see Supplementary Fig. 1b). Therefore, the barrier height and the built-in potential at the respective contacts needed to be considered separately, as follows[49,50]:

The 'left' contact (Au-OMeTP-MAPbI$_3$):

$$q\Phi_{Bn} = q(\Phi_m^{(a)} - \chi_{per}) = 4.5\,eV - 3.9\,eV = 0.6\,eV$$

$$qV_n = q(\Phi_s - \chi_{per}) = 4.7\,eV - 3.9\,eV = 0.8\,eV$$

$$qV_{bi}^{(a)} = q(\Phi_{Bn} - V_n) = q(\Phi_m^{(a)} - \Phi_s) = 0.6\,eV - 0.8\,eV = -0.2\,eV \quad \text{- 'downwards' MAPbI}_3$$
energy band bending.

or,

$$q\Phi_{Bp} = E_g - q(\Phi_m^{(a)} - \chi_{per}) = 1.6\,eV - 4.5\,eV + 3.9\,eV = 1\,eV$$

$$qV_p = E_g - qV_n = E_g - q(\Phi_s - \chi_{per}) = 1.6\,eV - 4.7\,eV + 3.9\,eV = 0.8\,eV$$

$$qV_{bi}^{(a)} = q(\Phi_{Bp} - V_p) = 1\,eV - 0.8\,eV = 0.2\,eV \quad \text{- 'downwards' MAPbI}_3 \text{ energy band bending.}$$

$$E_g = q(\Phi_{Bn} + \Phi_{Bp}) = 0.6\,eV + 1\,eV = 1.6\,eV$$

The 'right' contact (MAPbI$_3$-ClTP-Au):

$$q\Phi_{Bn} = q(\Phi_m^{(b)} - \chi_{per}) = 5.1\,eV - 3.9\,eV = 1.2\,eV$$

$$qV_n = q(\Phi_s - \chi_{per}) = 4.7\,eV - 3.9\,eV = 0.8\,eV$$

$$qV_{bi}^{(b)} = q(\Phi_{Bn} - V_n) = 1.2\,eV - 0.8\,eV = 0.4\,eV \text{ - 'upwards' MAPbI}_3 \text{ energy band bending.}$$

Thus, the above calculations suggest that at thermal equilibrium in the dark there is a downwards MAPbI$_3$ energy band bending at the Au-OMeTP-MAPbI$_3$ M-S interface with a 0.6 eV barrier height, $q\Phi_{Bn}$, and a −0.2 eV built-in potential, $V_{bi}^{(a)}$, whereas at the MAPbI$_3$-ClTP-Au M-S interface there is an upwards MAPbI$_3$ energy band bending with a 1.2 eV barrier height and a 0.4 eV built-in potential (see Supplementary Fig. 1b). The band diagram of the asymmetric 'a'-MAPbI$_3$-'b' device can be represented as front-to-back connected Schottky diodes.

**Theoretical PCE for Schottky Solar Cells.** The upper-limit to the maximum theoretical PCE of a cell consisting of a single metal-semiconductor contact (Schottky solar cell) was estimated based on simplified formalism of Pulfrey *et al.*[29] by taking into account the reverse current due to thermionic emission over the Schottky barrier in addition to the (short-circuit) photocurrent. The model assumes unit absorbance of all photons with energies above the semiconductor (perovskite) band gap ($E_g$ = 1.6 eV), and ignores all possible non-radiative recombination mechanisms.

The 'a'-MAPbI$_3$ contact in 'a'-MAPbI$_3$-'b' and 'a'-MAPbI$_3$-Au device constructs can be considered as an Ohmic contact since the work function of electrode 'a' (Au/OMeTP; 4.5 eV) is smaller than the work function of the perovskite (4.7 eV)[51]. Thus, this contact does not cause any barrier for electron flow from the semiconductor (MAPbI$_3$) to the electrode 'a'. Work functions of Au (4.9 eV) and electrode 'b' (Au/ClTP; 5.1 eV) are larger than the work function of the perovskite (4.7 eV). Therefore, MAPbI$_3$-'b' and MAPbI$_3$-Au contacts in the 'a'-MAPbI$_3$-Au and 'a'-MAPbI$_3$-'b' device constructs form the rectifying contacts. Hence, Schottky barriers only at the MAPbI$_3$-'b' (1.2 eV) and MAPbI$_3$-Au (1 eV) contacts need to be considered in determining the maximum PCE for the 'a'-MAPbI$_3$-'b' (~15%) and 'a'-MAPbI$_3$-Au (10%) devices. In the Au-MAPbI$_3$-'b' device construct the left (Au-MAPbI$_3$) and right (MAPbI$_3$-'b') contacts are both rectifying, therefore, the barrier heights for both contacts need to be considered to derive the total current due to thermionic emission over both of the rectifying barriers[52], and subsequently the PCE of the device.

**Data availability.** The authors declare that the main data supporting the findings of this study are available within the article and its Supplementary Information files. Extra data are available from the corresponding author upon request.

**Acknowledgments:** This work was performed in part at the Melbourne Centre for Nanofabrication (MCN) in the Victorian Node of the Australian National Fabrication Facility (ANFF). The authors thank Dr. Alexandr Simonov, Mr. Dijon Hoogeveen and Dr. Shannon A. Bonke for their help with electrochemistry experiments, as well as Mr. Soon Hock Ng and Mr. Julian Lloyd for their help with the design of the graphs. This work was funded through CSIRO Manufacturing as part of an Office of the Chief Executive Postdoctoral Fellowship (A. N. J.) and Science Leader Fellowship (U. B.). The authors also acknowledge the financial support from the Australian Renewable Energy Agency (ARENA) and the Australian Centre for Advanced Photovoltaics (ACAP).

**Author Contributions:** X.L., A.N.J. and U. B. designed and directed the study. A.R.P. and Y.C. deposited perovskite absorber layer. N.N.L., Y.Z. and Q.B. carried out the photocurrent and photoluminescence mapping measurements and analysis. D.E.G., N.W.D. carried out theoretical calculations. A.S.R.C. carried out Raman spectroscopy measurements and analysis. K.S. carried out the external quantum efficiency measurements and analysis. M.F. and L.S. performed the electrochemical desorption experiments. X.L., A.N.J. and U.B. wrote the manuscript. All authors commented on the paper.

**Author Information**: Reprints and permissions information is available at www.nature.com/reprints. The authors declare no competing financial interests. Readers are welcome to comment on the online version of the paper. Correspondence and requests for materials should be addressed to U.B. (udo.bach@monash.edu).